\documentclass[conference]{IEEEtran}

\usepackage{amsfonts,amsmath,amsthm}
\usepackage{graphicx,xspace,epsfig,syntonly,dsfont}
\usepackage{url,cite,bm,bbm}
\usepackage{algorithmicx}
\usepackage{balance}
\usepackage{subfigure}
\usepackage{stfloats}
\usepackage{diagbox}
\usepackage{array,makecell}
\usepackage{cleveref}
\usepackage[stable]{footmisc}

\newtheorem{remark}{Remark}

\long\def\symbolfootnote[#1]#2{\begingroup
	\def\thefootnote{\fnsymbol{footnote}}
	\footnote[#1]{#2}\endgroup}

\IEEEoverridecommandlockouts

\allowdisplaybreaks[4]

\newcommand{\mb}[1]{\mathbf{#1}}
\newcommand{\mcal}[1]{\mathcal{#1}}
\newcommand{\elem}[2]{\left(#1\right){#2}}
\newcommand{\stack}[2]{\pmb{\Xi}_{#2} \left( #1 \right)}
\newcommand{\ldef}{\stackrel{\Delta}{=}}
\newcommand{\real}[1]{\mathbb{R}^{#1}}
\newcommand{\load}{\text{L}}
\newcommand{\gen}{\text{Gen}}
\newcommand{\ess}{\text{ESS}}
\newcommand{\tth}[1]{{#1}^\text{th}}
\newcommand{\ps}[2]{{#1}^{\in#2}}
\newcommand{\ch}{{\text{ch}}}
\newcommand{\dis}{{\text{dis}}}
\newcommand{\bp}{\pmb{\phi}}
\newcommand{\init}{\text{init}}
\newcommand{\mw}{\text{MW}}
\newcommand{\mwh}{\text{MWh}}
\newcommand{\mvar}{\text{MVar}}
\newcommand{\cbrace}[1]{\left\{ #1 \right\}}

\title{\Large \bf Co-optimization of Battery Routing and Load Restoration for\\Microgrids with Mobile Energy Storage Systems
}
\author{
	\IEEEauthorblockN{Shourya Bose, Sifat Chowdhury, and Yu Zhang}
	\IEEEauthorblockA{\textit{Department of Electrical and Computer Engineering}}
	\IEEEauthorblockA{\textit{University of California, Santa Cruz}}
	\IEEEauthorblockA{Emails: \{\texttt{shbose,schowdh6,zhangy}\}@ucsc.edu}

}

\begin{document}
	\maketitle

	\begin{abstract}
		Mobile energy storage systems (MESS) offer great operational flexibility to enhance the resiliency of distribution systems in an emergency condition. The optimal placement and sizing of those units are pivotal for quickly restoring the curtailed loads. In this paper, we propose a model for load restoration in a microgrid while concurrently optimizing the MESS routes required for the same. The model is formulated as a mixed-integer second order cone program by considering the state of charge and evolution of the lower and upper bounds of battery capacities. 
		Simulation results tested on the IEEE 123-bus benchmark system demonstrate the efficacy of the proposed model.
	\end{abstract}

	\section{Introduction}\label{sect:intro}
	Extreme weather events induced by climate change severely affect the reliability of electric power systems. According to the U.S. Energy Information Administration, customers faced over eight hours of power interruption on average in 2020, which has doubled in just five years \cite{USEIA2021}. More frequent power outages due to high-impact low-probability (HILP) events result in tremendous financial loss, roughly \$150 billion per year \cite{USDOE2017}. In order to improve resiliency of the power system, different post contingency corrective actions can be adopted to effectively restore curtailed loads; e.g., deploying a microgrid (MG). A MG is a self-reliant energy system that is able to serve the loads within a boundary by its locally distributed energy resources even when the network is disconnected from the main grid. However, because of the inherent uncertainty associated with renewable sources and the capacity limit of the on-site generators, MGs are not always sufficient enough to restore all the loads during a prolonged outage. Therefore, it necessitates the allocation of other response resources such as portable generators and energy storage systems (ESS); e.g., grid-scale battery packs.
	
	ESS units are fast responding, flexible and versatile resources that can store and deliver energy per situation demand. Most ESS are stationary and installed along the permanent infrastructure of the network. However, they have finite capacity and need to be recharged once the storage capacity is fully exhausted which is not feasible during an outage. In this regard, mobile ESS (MESS) can be very helpful. MESSs are vehicle mounted standalone ESSs that can be integrated in prioritized locations from off-site to curb the additional load curtailments. This emerging technology is faster and more environment friendly compared with other emergency relief alternatives; e.g., portable generators and operating reserves. Unlike topology switching and adaptive MGs, they do not require advanced communication infrastructure and can operate in a plug-and-play manner. However, it is important to adopt a proper operation strategy for the optimal placement and sizing of MESS to enhance the system resilience.

	The potential challenges of optimizing the MESS integration in a distribution system (DS) have received wide attention. Existing approaches include stochastic optimization \cite{stochastic}, robust optimization \cite{robust}, as well as a combination of both \cite{R&S_combo}. Authors of \cite{Kim} develop a two-stage stochastic mixed integer second order cone programming (MISOCP) model for the economic operation of a MG equipped with MESS during normal and emergency conditions. Another type of two-stage optimum allocation strategy has been carried out where the MESSs are pre-positioned prior to the event and are dynamically dispatched during the following contingency period \cite{robust,two_step}. \cite{RES_uncertainty} incorporates the uncertainty associated with renewable generation in the load restoration (LR) problem by formulating the routing and scheduling of MESS as a chance constraint program. In addition, the uncertainty due to the transportation of the MESSs has also been considered for developing the optimum route planning model \cite{transport_uncertainty_1,transport_uncertainty_2}. 
	
	All previous works have considered a fixed upper and lower bound of the state of charge (SoC) for the ESS units. Treating these quantities as variables allows for incorporation of constraints which model a MESS being transported from one bus to the other. The main contribution of this paper is to propose an LR problem involving evolution of the SoC bounds as decision variables. Dynamic changes of these bounds occur as a consequence of optimal routing of MESSs between different buses.
	
	\textit{Notation}: The symbols $\real{}$ and $\mathbb{N}_0$ denote the set of real numbers and non-negative integers, respectively. Vectors are represented in boldface. $\elem{\mb{a}}{j}$ is the $\tth{j}$ element of $\mb{a}$. Symbols $\mb{0}$ and $\mb{1}$ denote vectors of all zeros and ones. For multiple sequences of vectors $\{ \mb{a}^{(1)}_k \}_{k\in\mcal{K}},\cdots,\{\mb{a}^{(C)}_k \}_{k\in\mcal{K}} $, the expression $\stack{\mb{a}^{(1)}_k,\cdots,\mb{a}^{(C)}_k}{\mcal{K}}$ represents the stacked vector $\left[{\mb{a}^{(1)}_{1}}^\top,\cdots,{\mb{a}^{(1)}_{|\mcal{K}|}}^\top,\cdots,{\mb{a}^{(C)}_{1}}^\top,\cdots,{\mb{a}^{(C)}_{|\mcal{K}|}}^\top\right]^\top$. $[N]$ denotes the set $\{ 1,\cdots,N \}$.
	For $\mb{a}\in\real{d}$ and $\mcal{D} \subseteq [d]$, we let $\ps{\mb{a}}{\mcal{D}}$ denote the subvector of $\mb{a}$ corresponding to the indices in $\mcal{D}$. Element-wise vector multiplication is denoted by $\odot$. The abbreviation u.a.r. stands for `uniformly at random'.

	\section{Problem Formulation} \label{sec:pf}
	A MG can be represented by a directed tree graph $\mcal{G}\ldef (\mcal{N},\mcal{E})$, where $\mcal{N}$ denotes the set of buses and $\mcal{E}$ denotes the set of branches. We assume that the buses can be either load buses, generator buses, or ESS buses, i.e. $\mcal{N} = \mcal{N}^\load \cup \mcal{N}^\gen \cup \mcal{N}^\ess$. We consider the problem over a time horizon of $[T]$. Let $N \ldef |\mcal{N}|$ and $E\ldef |\mcal{E}|$. Further, let $N_\load \ldef |\mcal{N}^\load|$, $N_\gen \ldef |\mcal{N}^\gen|$, and $N_\ess \ldef |\mcal{N}^\ess|$. In general, renewable sources with curtailment capability may also be included in the MG using a generation forecast as shown in~\cite{SB-YZ:2021}. For sake of brevity, and in order to demonstrate the impact of MESSs, we do not consider renewable sources in the MG.
	\subsection{Objective function}
	The optimal resource allocation task is to maximize the load restoration (LR) while minimizing the cost of power generation and MESS transportation over a time horizon $[T]$. The objective function is given as
	\begin{gather}
		J = \sum_{t= 1}^T \left( ({\pmb{\zeta}^{\mb{r}}_t})^\top \mb{r}_t  - ({\pmb{\zeta}^{\gen}_t})^\top \mb{p}^\gen_t \right) - \sum_{\nu = 1}^M \sum_{\omega=1}^H (\pmb{\zeta}^\mb{k}_{\nu,\omega})^\top  \mb{k}^\omega_\nu,
		\label{obj_func}
	\end{gather}
where the load pickup, active power by the generators and the number of MESS units are collected by the vectors $\mb{r}_t$, $\mb{p}_t$ and $\mb{k}_\nu^\omega$, respectively. The coefficients $\pmb{\zeta}^{\mb{r}}_t$, $\pmb{\zeta}^{\gen}_t$ and $\pmb{\zeta}^\mb{k}_{\nu,\omega}$ are the incentives for load pickup, power generation cost and MESS transportation cost, respectively. The decision variables will be explained at length in the following sections.

	\subsection{Constraints\footnote{Unless specified otherwise,  constraints described in this subsection hold for all $t \in [T]$.}}
	
	\subsubsection*{Power flow constraints} Let $\mb{v}_t\in\real{N}$, $\mb{p}_t\in\real{N}$, and $\mb{q}_t\in\real{N}$ denote the vectors of bus voltage magnitude squared, bus real power injection, and bus reactive power injection respectively.  Let $\mb{l}_t\in\real{E}$, $\mb{P}_t\in\real{E}$, and $\mb{Q}_t\in\real{E}$ denote the vector of branch current magnitude squared, branch real power flow, and branch reactive power flow respectively. The SOCP relaxation to DistFlow equations is given as \cite{Distflow}:
	\begin{gather}
		\sum_{\substack{k\in\mcal{N}:e=\\(j,k)\in\mcal{E}}} \elem{\mb{P}_t}{e} = \elem{\mb{p}_t}{j} + \sum_{\substack{i\in\mcal{N}:e'=\\(i,j)\in\mcal{E}}} \elem{\mb{P}_t}{e'} - r_{e'}\elem{\mb{l}_t}{e'} \label{real_power} \\
		\sum_{\substack{k\in\mcal{N}:e=\\(j,k)\in\mcal{E}}} \elem{\mb{Q}_t}{e} = \elem{\mb{q}_t}{j} + \sum_{\substack{i\in\mcal{N}:e'=\\(i,j)\in\mcal{E}}} \elem{\mb{Q}_t}{e'} - x_{e'}\elem{\mb{l}_t}{e'} \label{reactive_power} \\
		\elem{\mb{v}_t}{i} - \elem{\mb{v}_t}{j} = 2(r_e\elem{\mb{P}_t}{e}+x_e\elem{\mb{Q}_t}{e}) - |z_e|^2\elem{\mb{l}_t}{e} \label{voltage} \\
		\elem{\mb{v}_t}{i} \elem{\mb{l}_t}{e} \geq \elem{\mb{P}_t}{e}^2 + \elem{\mb{Q}_t}{e}^2 \label{squared_quantities}
	\end{gather}
	where  \eqref{real_power}-\eqref{reactive_power} hold for all $j \in \mcal{N}$, $z_e \ldef r_e+jx_e$ denotes the impedance of branch $e$, and \eqref{voltage}-\eqref{squared_quantities} hold for all $e = (i,j) \in \mcal{E}$. \\
	
	\subsubsection*{Voltage constraints} To maintain the nominal operation, the voltage at each bus ($\forall j \in \mcal{N}$) and current through each branch ($\forall e = (i,j)\in \mcal{E}$)  are restricted within a permissible range according to \eqref{volt_bound} and \eqref{current_bound}. Furthermore, constraint~\eqref{volt_bound} ensures a unity voltage squared magnitude for the slack bus.
	\begin{gather}
		\underline{\mb{v}} \leq \mb{v}_t \leq \bar{\mb{v}}, \;\; \elem{\mb{v}_t}{\text{slackbus}} = 1 \label{volt_bound} \\
		\mb{0} \leq \mb{l}_t \leq \bar{\mb{l}}  \label{current_bound}
	\end{gather}
	
	\subsubsection*{Generator constraints} 
	Let $\mb{p}_t^\load \ldef \ps{\mb{p}_t}{\mcal{N}^\load}$, $\mb{p}_t^\gen \ldef \ps{\mb{p}_t}{\mcal{N}^\gen}$, and $\mb{p}_t^\ess \ldef \ps{\mb{p}_t}{\mcal{N}^\ess} $ (similarly $\mb{q}_t^\load\ldef \ps{\mb{q}_t}{\mcal{N}^\load}$, $\mb{q}_t^\gen\ldef \ps{\mb{q}_t}{\mcal{N}^\gen}$, and $\mb{q}_t^\ess \ldef \ps{\mb{q}_t}{\mcal{N}^\ess}$) denote the sub-vectors of $\mb{p}_t$ (similarly $\mb{q}_t$) representing the real (similarly reactive) power injections from load, generator, and ESS buses respectively. Constraints \eqref{fuel_burn}-\eqref{gen_limit} which hold for all $j \in [N_\gen]$ represent the generator dynamics. Let $\mb{f}_t\in\real{N_\gen}$ be the amount of fuel remaining at time $t$ and $\tau_j$ represent the conversion efficiency between fuel expenditure and power production, \eqref{fuel_burn} explains the evolution of fuel and sets the initial amount of fuel, and \eqref{fuel_nonnegativity} ensures the nonnegativity characteristics of $\mb{f}_t$. \eqref{ru_and_rd}-\eqref{gen_limit} fix the ramp up/down and generation limit. 
	\begin{gather}
		\elem{\mb{f}_{t+1}}{j} = \elem{\mb{f}_{t}}{j} - \tau_j \Delta {t} \elem{\mb{p}^\gen_t}{j} \label{fuel_burn}, \; \elem{\mb{f}_1}{j} = F_j\\
		\mb{f}_t \geq \mb{0}, \forall t \in [T+1] \label{fuel_nonnegativity} \\
		P_j^\downarrow \leq \elem{\mb{p}^\gen_{t+1}}{j} -  \elem{\mb{p}^\gen_{t}}{j} \leq P_j^\uparrow, \; \forall t \in [T-1]\label{ru_and_rd}\\
		\elem{\mb{p}^\gen}{j} \leq P_j^\uparrow \label{init_ru}\\
		\mb{0} \leq \mb{p}^\gen_t \leq \bar{\mb{p}}^\gen. \label{gen_limit}
	\end{gather}
	\subsubsection*{Reactive power constraints} We assume that the ESS and generator buses are equipped with controllable inverters. Hence, the reactive power bounds are given as follows:
	\begin{gather}
		\elem{\underline{\mb{q}_t}}{j} \leq \elem{{\mb{q}_t}}{j} \leq \elem{\bar{\mb{q}_t}}{j}, \; \forall j \in \mcal{N}^\gen \cup \mcal{N}^\ess.
	\end{gather}
	
	\subsubsection*{Load pickup constraints} We assume to have a forecast for the real and reactive power demand respectively denoted as $\hat{\mb{p}}_t^\load$ and $\hat{\mb{q}}_t^\load$ for all load buses i.e. $j \in [N_\load]$, for all $t \in [T]$ and having a constant power factor, then the constraints  for the \emph{pickup vector} $\mb{r}_t \in \real{N_\load}$ follow \eqref{r}. We also assume monotonic LR i.e. not to drop a load once it is picked up which is enforced in \eqref{monotonic_load_pickup}.
	\begin{gather}
		\elem{\mb{r}_t}{j} = \frac{\elem{\mb{p}_t^\load}{j}}{\elem{\hat{\mb{p}}_t^\load}{j}} = \frac{\elem{\mb{q}_t^\load}{j}}{\elem{\hat{\mb{q}}_t^\load}{j}}, \;\; \mb{0} \leq \mb{r}_t \leq \mb{1} \label{r}\\
		\mb{r}_t \leq \mb{r}_{t+1}, \; \forall t \in [T-1] \label{monotonic_load_pickup}.
	\end{gather}
	
	\subsubsection*{MESS constraints} This paragraph contains the most important set of constraints from \eqref{comp_charging}--\eqref{UB_S}, regarding the ESS which hold for all $j \in [N_\ess]$. Let $\mb{p}^\ch_t$ and $\mb{p}^\dis_t$ be the charging and discharging power of the ESS units, $\eta^\text{ch}$ and $\eta^\text{dis}$ be the corresponding efficiency and $\mb{d}_t$ be the binary decision variable indicating the charging or discharging state. Then the complimentarity constraint is specified in \eqref{comp_charging} and \eqref{comp_discharging}.~\eqref{power_balance_ess} denotes the power balance at each ESS.
	\begin{gather}
		\mb{0} \leq \mb{p}^\ch_t \leq \mb{d}_t \odot \bar{\mb{p}}^\ch \label{comp_charging}\\
		\mb{0} \leq \mb{p}^\dis_t \leq (\mb{1}-\mb{d}_t) \odot \bar{\mb{p}}^\dis \label{comp_discharging}\\
		\mb{p}_t^\ess = -\mb{p}^\ch_t + \mb{p}^\dis_t\label{power_balance_ess}.
	\end{gather}
	
	Now we model the evolution of state of charge (SoC) along with its flow on the transportation network. Let $\mb{s}_t$ and $\mb{S}_t$ denote the SoC of the stationary and mobile ESS units respectively. $\bp_{t}$ indicates the additional SoC gained (lost) due to the inclusion (removal) of MESS units as expressed in \eqref{phi_t}. $\nu$ stands for different types of MESS (according to their capacity) and $R_{j,\nu}$ denotes the number of MESSs of type $\nu$ connected initially at bus j. 
	The routing of MESS will be done over dynamic sets of all feasible routes \emph{viz.} ${\mcal{E}_{\omega,t}} \subseteq \mcal{N}^\ess \times \mcal{N}^\ess$ defined $\forall t \in [T-\omega], \forall \omega \in [H]$, where $\omega$ denotes the necessary time for transporting a batch of MESSs from one bus to another. We assume the maximum transport time $H$ is known and that $H<T$. Thus any MESS traveling on $\mcal{E}_{\omega,t}$ departs at $t$ and arrives at its destination at $t+\omega$. On each time step, we have $H$ transport graphs denoting feasible paths to get MESSs delivered to their ideal destinations in 1 through $H$ time steps respectively.
	$\mb{S}_{t,\nu}^{\omega}$ and $\mb{k}^\omega_{t,\nu}$ respectively represent the vector of total SoC and the number of travelling units of MESS type $\nu$ which were sent off on time step $t$ and will reach their destination $\omega$ time steps later. $\bar{s}^*_{\nu}$ and $\underline{s}^*_{\nu}$ are respectively the upper and lower bound of the SoC of each MESS type $\nu$. \eqref{SoC_t}--\eqref{phi_t} define  the SoC evolution of each ESS bus at any time $t$. The evolution of the lower bound (LB) of these SoCs are determined in \eqref{LB_SoC}--\eqref{LB_SoC_last} and the upper bounds (UB) in \eqref{UB_SoC}--\eqref{UB_SoC_last}. 
	
	The main idea is that a traveling MESS unit can pick any amount of energy or SoC from the existing set of MESSs at the ESS bus as long as their upper and lower SoC capacity bounds are not violated. We ignore any transient effects arising from the integration and detachment of MESS to and from ESS buses. When the MESSs are being combined at another ESS bus, their SoC and associated UB and LB are also additive. These cumulative quantities are modeled in the following set of equations.
	\begin{multline}
		\elem{\mb{s}_{t+1}}{j} = \elem{\mb{s}_t}{j} + \elem{\bp_{t}}{j}  + \eta^\text{ch}\Delta t \elem{\mb{p}^\ch_t}{j} - \frac{\Delta t}{\eta^\dis}  \elem{\mb{p}^\dis_t}{j} \label{SoC_t}
	\end{multline}
	\begin{gather}
		\elem{\mb{s}_1}{j} = \sum_{\nu=1}^{M} R_{j,\nu} \sum_{r=1}^{R_{j,\nu}} s^\init_{r,j,\nu}  \label{SoC_init} \\
		\underline{\mb{s}}_t \leq \mb{s}_t + \bp_t \leq \bar{\mb{s}}_t, \; \forall t \in [T+1] \label{SoC_t_limit}
	\end{gather}
	\begin{align}
		\elem{\bp_t}{j} &= \sum_{\substack{t' =\\ \max\{t-H,1 \}}}^{t-1} \sum_{\substack{e =(i,j)\\\in\mcal{E}_{t-t',\nu}}} \sum_{\nu=1}^M \elem{{\mb{S}}^{t-t'}_{t',\nu}}{e} \nonumber \\&- \sum_{t'=t+1}^{\min\{t+H,T\}} \sum_{\substack{e =(j,k)\\\in\mcal{E}_{t-t',\nu}}} \sum_{\nu=1}^M \elem{{\mb{S}}^{t-t'}_{t',\nu}}{e}; \quad \bp_{T+1} = \mb{0} \label{phi_t}
	\end{align}
	\begin{align}
		\elem{\underline{\mb{s}}_{t+1}}{j} = \elem{\underline{\mb{s}}_{t}}{j} &+ \sum_{\substack{t' =\\ \max\{t-H,1 \}}}^{t-1} \sum_{\substack{e =(i,j)\\\in\mcal{E}_{t'-t,\nu}}} \sum_{\nu=1}^M \elem{\underline{\mb{S}}^{t-t'}_{t',\nu}}{e} \nonumber \\&-  \sum_{t'=t+1}^{\min\{t+H,T\}} \sum_{\substack{e =(j,k)\\\in\mcal{E}_{t'-t,\nu}}} \sum_{\nu=1}^M \elem{\underline{\mb{S}}^{t-t'}_{t',\nu}}{e} \label{LB_SoC}
	\end{align}
	\begin{gather}
		\elem{\underline{\mb{s}}_1}{j} = \sum_{\nu=1}^{M} R_{j,\nu} \underline{s}^*_{\nu}; \quad \elem{\underline{\mb{s}}_{T+1}}{j} = \elem{\underline{\mb{s}}_{T}}{j} \label{LB_SoC_last}
	\end{gather}
	\begin{align}
		\elem{\bar{\mb{s}}_{t+1}}{j} = \elem{\bar{\mb{s}}_{t}}{j} &+ \sum_{\substack{t' =\\ \max\{t-H,1 \}}}^{t-1} \sum_{\substack{e =(i,j)\\\in\mcal{E}_{t-t',\nu}}} \sum_{\nu=1}^M \elem{\bar{\mb{S}}^{t-t'}_{t',\nu}}{e} \nonumber \\&- \sum_{t'=t+1}^{\min\{t+H,T\}} \sum_{\substack{e =(j,k)\\\in\mcal{E}_{t-t',\nu}}} \sum_{\nu=1}^M \elem{\bar{\mb{S}}^{t-t'}_{t',\nu}}{e}  \label{UB_SoC}
	\end{align}
	\begin{gather}
		\elem{\bar{\mb{s}}_1}{j} = \sum_{\nu=1}^{M} R_{j,\nu} \bar{s}^*_{\nu} ; \quad \elem{\bar{\mb{s}}_{T+1}}{j} = \elem{\bar{\mb{s}}_{T}}{j} \label{UB_SoC_last}
	\end{gather}
	\begin{align}
		&\sum_{j\in[N_\ess]} \elem{\underline{\mb{s}_t}}{j} + \sum_{\substack{t' = \\ \max\{ t-H-1 ,1\}}}^{t-1} \sum_{\substack{e = (i,j)\\ \in \mcal{E}_{t'-t,\nu}}} \sum_{\nu=1}^M \elem{\underline{\mb{S}}^{t-t'}_{t',\nu}}{e} \nonumber \\&+ \sum_{\substack{t' = t+1}}^{\substack{\min\{t+H,\\T\}}} \sum_{\substack{e = (j,k)\\ \in \mcal{E}_{t'-t,\nu}}} \sum_{\nu=1}^M \elem{\underline{\mb{S}}^{t-t'}_{t',\nu}}{e} = \sum_{j\in[N_\ess]}\sum_{\nu=1}^{M} R_{j,\nu} \underline{s}^*_{\nu} \label{LB_total}
	\end{align}
	\begin{align}
		&\sum_{j\in[N_\ess]} \elem{\bar{\mb{s}_t}}{j} + \sum_{\substack{t' = \\ \max\{ t-H-1 ,1\}}}^{t-1} \sum_{\substack{e = (i,j)\\ \in \mcal{E}_{t'-t,\nu}}} \sum_{\nu=1}^M \elem{\bar{\mb{S}}^{t-t'}_{t',\nu}}{e} \nonumber \\&+ \sum_{\substack{t' = t+1}}^{\substack{\min\{t+H,\\T\}}} \sum_{\substack{e = (j,k)\\ \in \mcal{E}_{t'-t,\nu}}} \sum_{\nu=1}^M \elem{\bar{\mb{S}}^{t-t'}_{t',\nu}}{e} = \sum_{j\in[N_\ess]}\sum_{\nu=1}^{M} R_{j,\nu} \bar{s}^*_{\nu} \label{UB_total}
	\end{align}
	\begin{gather}
		\underline{\mb{S}}^\omega_{t,\nu} = \mb{k}^\omega_{t,\nu} \underline{s}_\nu^*, \; \forall \nu \in [M], \forall \omega \in [H],\forall t \in [T-\omega] \label{LB_S}\\
		\bar{\mb{S}}^\omega_{t,\nu} = \mb{k}^\omega_{t,\nu} \bar{s}_\nu^*, \; \forall \nu \in [M], \forall \omega \in [H], \forall t \in [T-\omega] \label{UB_S}.
	\end{gather}
	
	Equation \eqref{SoC_init} determines the initial SoC of any bus $j$ as the total SoC contributed by all units of type $\nu$, summed over all values of $\nu$. \eqref{SoC_t_limit} limits SoC at any time within its upper and lower bound. In \eqref{phi_t}, $\elem{\bp_t}{j}$ is the net SoC contributed by the mobile units present at bus $j$, at time $t$ and is expressed as the difference between the summation of the SoC of all incoming MESS units those are on the way to reach the same bus at the same time and the units which are going to leave. It also ensures that the transportation scenario will only last till $T$. \eqref{LB_SoC} defines the dynamic lower bound of the SoC for any bus $j$ and \eqref{UB_SoC} defines the dynamic upper bound. \eqref{LB_SoC_last} and \eqref{UB_SoC_last} fix these two quantities for the boundary time steps, with $R_{j,\nu}$ denoting the initial number of units of type $\nu$ at bus $j$. \eqref{LB_total} controls the LB of SoC provided by all the units (both static and dynamic) across the entire set of ESS buses at any time $t$ in such a way that it must be equal to the total LB of all units of type $\nu$ present there at the initial time step. \eqref{UB_total} does the same for the UB. Finally, \eqref{LB_S} and \eqref{UB_S} denote the LB and UB of the SoC currently on the route to be equal to a constant multiple of the LB and UB of each travelling unit of the corresponding type $\nu$.

	\subsubsection*{Discrete constraints} There are two sets of discrete constraints where the first one in \eqref{binary_d} is regarding the complementarity conditions for charging and discharging batteries. The second one in \eqref{discretized_MESS} only allows integer number of travelling ESS. Finally, we define the stacked variable in \eqref{stacked_k} that is used in the objective \eqref{obj_func}.
	\begin{gather}
		\mb{d}_t \in \{ 0,1 \}^{N_\ess} \label{binary_d} \\
		\mb{k}^\omega_{t,\nu} \in \mathbb{N}_0, \; \forall \nu \in [M], \forall \omega \in [H], \forall t \in [T-\omega] \label{discretized_MESS}\\
		\mb{k}^\omega_\nu \ldef \stack{\mb{k}^\omega_{t,\nu}}{[T-\omega]}, \; \forall \omega \in [H], \forall \nu \in [M] \label{stacked_k}.
	\end{gather}
	
\begin{remark}
In practice, vehicles used for transporting MESS units have finite transport capacities, which can be modeled using box constraints on the number of traveling MESSs. In our formulation, we have explicitly modelled the number of mobile units using $\mb{k}^\omega_\nu$. This variable $\mb{k}^\omega_\nu$ may be placed inside any arbitrary box constraint. Due to space limitation, we skip said constraints in the current work.
\end{remark}

The following remark highlights that both stationary and mobile ESS can coexist in a MG.
\begin{remark}
    Each ESS bus may have a certain quantity of stationary ESSs attached to it, in addition to all the MESSs that may arrive and leave over $[T]$. In the current formulation, we assume that the entire capacity at each ESS bus is contributed by MESSs (\eqref{SoC_init}, \eqref{LB_SoC_last}, and~\eqref{UB_SoC_last}). However, additional linear constraints on $\mb{s}_t$, $\bar{\mb{s}}_t$ and $\underline{\mb{s}}_t$ can be formulated to incorporate stationary ESS units at a given ESS bus.
\end{remark}
	
	\subsubsection*{Co-optimization problem} To this end, our formulated optimization problem is an MISOCP, which aims at maximizing the objective~\eqref{obj_func}, subject to the constraints~\eqref{real_power}--\eqref{stacked_k}. The decision 
	variables are 
	$\stack{\mb{v}_t,\mb{p}_t,\mb{q}_t,\mb{l}_t,\mb{P}_t,\mb{Q}_t}{[T]}$, $\stack{\mb{p}^\ch_t,\mb{p}^\dis_t,\mb{r}_t,\mb{d}_t}{[T]}$, $\stack{\mb{f}_t,\mb{s}_t,\bar{\mb{s}}_t,\underline{\mb{s}}_t,\pmb{\phi}_t}{[T+1]}$, and 
	$\cbrace{\stack{{\mb{S}}^\omega_{t,\nu},\bar{\mb{S}}^\omega_{t,\nu},
	\underline{\mb{S}}^\omega_{t,\nu},{\mb{k}}^\omega_{t,\nu}}{[T-\omega]}}_{\omega\in[H],\nu\in[M]}$\,.
	

	\vskip 0.25in
	\section{Simulation Results} \label{sec:tests}
	In this section, we present simulation results to demonstrate the advantages of our proposed LR formulation and method. As a baseline, we compare the performance of the proposed method with the LR with static ESSs (i.e. no routing of MESS). We use the IEEE 123 bus distribution system to emulate the MG. This system prescribes MG network topology and load bus locations, as well as their demanded real and reactive power. The system has six switches that are all set to be open for the purpose of simulation. This partitions the network into three independent MGs, and therefore making the system suitable for demonstrating advantages of MESSs. 
    The parameter values are listed in Table~\ref{fig:simulation_parameters}.
	
	\begin{table}[!tb]
		\centering
		\caption{Simulation parameters}
		\resizebox{\columnwidth}{!}{
		\fontsize{7}{8}\selectfont
			\begin{tabular}{|c|c|}
				\hline
				\textbf{Parameter} & \textbf{Value}\\
				\hline\hline
				$N,E,N_\ess,N_\gen,N_\load,T$ & 123, 118, 19, 19, 85, 10\\
				\hline
				$\mcal{N}^\ess$ & \makecell{$\{ 3,8,18,21,23,25,54,57,61,67,72,$\\$89,91,93,115,116,120,121,122 \}$}\\
				\hline
				$\mcal{N}^\gen$ & \makecell{$\{ 13,14,15,26,27,36,40,44,78,81,97,$\\$101,105,108,110,117,118,119,123\}$}\\
				\hline
				$\mcal{N}^\load$ & $[123]\setminus \mcal{N}^\ess \cup \mcal{N}^\gen$\\
				\hline
				$\underline{v},\bar{v},l$ & 0.95, 1.05, $\infty$ (for all buses \& lines)\\
				\hline
				$P^\uparrow_j, P^\downarrow_j, \bar{P}^\gen_j, \tau, \underline{q}, \bar{q}$ & \makecell{$0.15\mw~\forall j, -0.1\mw~\forall j, 0.46\mw~\forall j,$\\$1,-0.24~\mvar,  0.24~\mvar$}\\
				\hline
				$M, R_{j,1}, R_{j,2},N_E$ & $2, 10 \forall j, 20 \forall j,40$\\
				\hline
				$\underbar{s}^*_1, \underbar{s}^*_2,\bar{s}_1^*,\bar{s}_2^*$ & $ 0.025\mwh, 0.005\mwh, 0.25\mwh,0.05\mwh$\\
				\hline
				$\mcal{E}_{\omega,T}$ & \makecell{ $\{(121,61),(121,72),(121,120),(57,120),$\\
					$(121,116),(21,67),(21,122),(23,121),$\\
					$(25,89),(54,121),(54,25),(54,115),$\\$(54,116),(54,120),(57,54),(61,116),$\\
					$(67,23),(67,116),(67,120),(72,25),$\\$(72,57),(72,89),(72,93),(89,67),$\\
					$(89,72),(89,122),(91,115),(91,120),$\\$(93,121),(93,89),(115,57),(116,72),$\\
					$(116,120),(120,57),(120,89),(121,72),$\\$(121,93),(122,121),(122,61),(122,112)\}$\\$\forall \omega \in [M], \forall t\in[T-\omega]$}\\
				\hline
				$\bar{p}^\ch_j, \bar{p}^\dis_j, \Delta t, \eta^\ch, \eta^\dis$ & $0.16\mw~\forall j, 0.16\mw~\forall j, 1h, 0.9, 0.9$\\
				\hline
				$s^\init_{r,j,\nu}, F_j$ & $\in[0.5\bar{s}^*_\nu, 0.65\bar{s}^*_\nu] \text{ u.a.r. }\forall r, \forall j, 3.55\text{ units }\forall j$\\ 
				\hline
		\end{tabular}}
		\label{fig:simulation_parameters}
	\end{table}
	To make the comparison between two scenarios meaningful, we start with all MESSs fixed to their respective initial ESS buses at time $t=1$. In case of LR without MESSs, they are kept static at their initial locations. Otherwise, they are allowed to be routed between the ESS buses. We let the maximum window of delivery be $H=3$ for the transportation graph $\mcal{E}_{\omega,t}$. We also fix the initial values for fuel and SoC and load demand for both scenarios, so that any possible increase in restored load stems directly from the routing of MESSs. 
	\begin{table*}[!t]
		\footnotesize
		\parbox[t][][t]{.33\linewidth}{\centering\begin{tabular}{|c|c|c|c|}
				\hline
				\textbf{From} (bus) & \textbf{To} (bus) & \makecell{\textbf{ESS}\\\textbf{Type $\nu$}} & \textbf{Qty.}\\
				\hline\hline
				25 ($t = 2$) & 89 ($t = 3$) & 1 & 4\\
				\hline
				54 ($t = 2$) & 3 ($t = 3$) & 1 & 1\\
				\hline
				57 ($t = 2$) & 54 ($t = 3$) & 1 & 3\\
				\hline
				61 ($t = 2$) & 116 ($t = 3$) & 1 & 1\\
				\hline
				21 ($t = 3$) & 122 ($t = 4$) & 1 & 1\\
				\hline
				120 ($t = 3$) & 57 ($t = 4$) & 1 & 3\\
				\hline
				3 ($t = 4$) & 120 ($t = 5$) & 1 & 1\\
				\hline
				23 ($t = 4$) & 3 ($t = 5$) & 1 & 4\\
				\hline
				54 ($t = 4$) & 3 ($t = 5$) & 1 & 2\\
				\hline
				61 ($t = 4$) & 116 ($t = 5$) & 1 & 2\\
				\hline
				89 ($t = 4$) & 67 ($t = 5$) & 1 & 1\\
				\hline
				72 ($t = 5$) & 93 ($t = 6$) & 1 & 2\\
				\hline
		\end{tabular}}
		\hfill
		\parbox[t][][t]{.33\linewidth}{\centering\begin{tabular}{|c|c|c|c|}
				\hline
				\textbf{From} (bus) & \textbf{To} (bus) & \makecell{\textbf{ESS}\\\textbf{Type $\nu$}} & \textbf{Qty.}\\
				\hline\hline
				120 ($t = 5$) & 57 ($t = 6$) & 1 & 3\\
				\hline
				122 ($t = 5$) & 116($t = 6$) & 1 & 1\\
				\hline
				21 ($t = 9$) & 122 ($t = 10$) & 1 & 2\\
				\hline
				122 ($t = 1$) & 116 ($t = 3$) & 1 & 7\\
				\hline
				72 ($t = 2$) & 93 ($t = 4$) & 1 & 1\\
				\hline
				72 ($t = 3$) & 93 ($t = 5$) & 1 & 2\\
				\hline
				21 ($t = 3$) & 67 ($t = 6$) & 1 & 6\\
				\hline
				116 ($t = 2$) & 72 ($t =3$) & 2 & 1\\
				\hline
				116 ($t = 3$) & 72 ($t = 4$) & 2 & 6\\
				\hline
				57 ($t = 4$) & 54 ($t = 5$) & 2 & 1\\
				\hline
				120 ($t = 4$) & 57 ($t = 5$) & 2 & 2\\
				\hline
				122 ($t = 4$) & 116 ($t = 5$) & 2 & 1\\
				\hline
		\end{tabular}}
		\hfill
		\parbox[t][][t]{.33\linewidth}{\centering\begin{tabular}{|c|c|c|c|}
				\hline
				\textbf{From} (bus) & \textbf{To} (bus) & \makecell{\textbf{ESS}\\\textbf{Type $\nu$}} & \textbf{Qty.}\\
				\hline\hline
				120 ($t = 6$) & 89 ($t = 7$) & 2 & 2\\
				\hline
				122 ($t = 6$) & 18 ($t = 7$) & 2 & 1\\
				\hline
				116 ($t = 9$) & 120 ($t = 10$) & 2 & 2\\
				\hline
				54 ($t = 1$) & 115 ($t = 3$) & 2 & 7\\
				\hline
				72 ($t = 5$) & 93 ($t = 7$) & 2 & 2\\
				\hline
				89 ($t = 1$) & 72 ($t = 4$) & 2 & 3\\
				\hline
				120 ($t = 1$) & 89 ($t = 4$) & 2 & 5\\
				\hline
				8 ($t = 2$) & 120 ($t = 5$) & 2 & 2\\
				\hline
				116 ($t = 2$) & 72 ($t = 5$) & 2 & 4\\
				\hline
				116 ($t = 2$) & 120 ($t = 5$) & 2 & 2\\
				\hline
				116 ($t =4$) & 120 ($t = 7$) & 2 & 1\\
				\hline
		\end{tabular}}
		\vskip 0.04in
		\caption{Optimal routing schedule. The departure and arrival time steps of MESSs are mentioned in parentheses.}
		\label{fig:optimal_routing}
	\end{table*}
	We design the weight vectors $\pmb{\zeta}^\mb{r}_t$ and $\pmb{\zeta}^\gen_t$ in the objective function~\eqref{obj_func} such that the cost of restoring one unit of load power is the same at all buses, while the cost of generating one unit of power is the same at any generator bus and in per unit terms is $80\%$ the cost of restoring one unit of load power. We also assign transportation costs (for one time step delivery) per MWh of $\bar{\mb{S}}$ to be $10\%$ the consumption cost of 1 MWh power, and every additional time step taken for routing adds $10\%$ to the transportation cost.
	\begin{figure}[!tb]
		\centering
		\includegraphics[width=0.7\linewidth,height=0.33\linewidth]{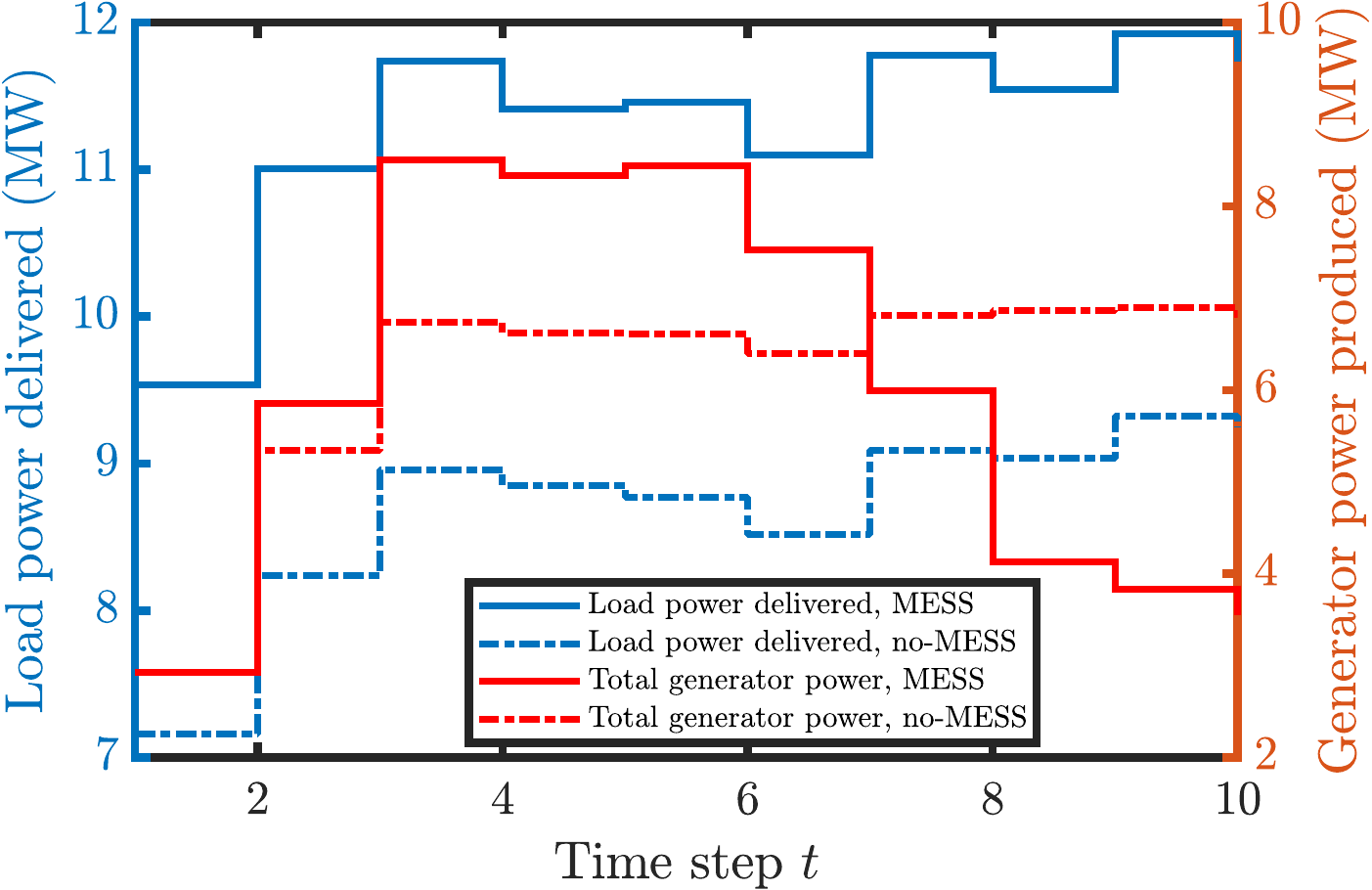}
		\caption{Comparison of load power restored and total generation power at different time steps achieved with and without MESS.}
		\label{fig:comparison_gen_load}
	\end{figure}
	
	\begin{figure}[!tb]
		\centering
		\includegraphics[width=0.7\linewidth,height=0.33\linewidth]{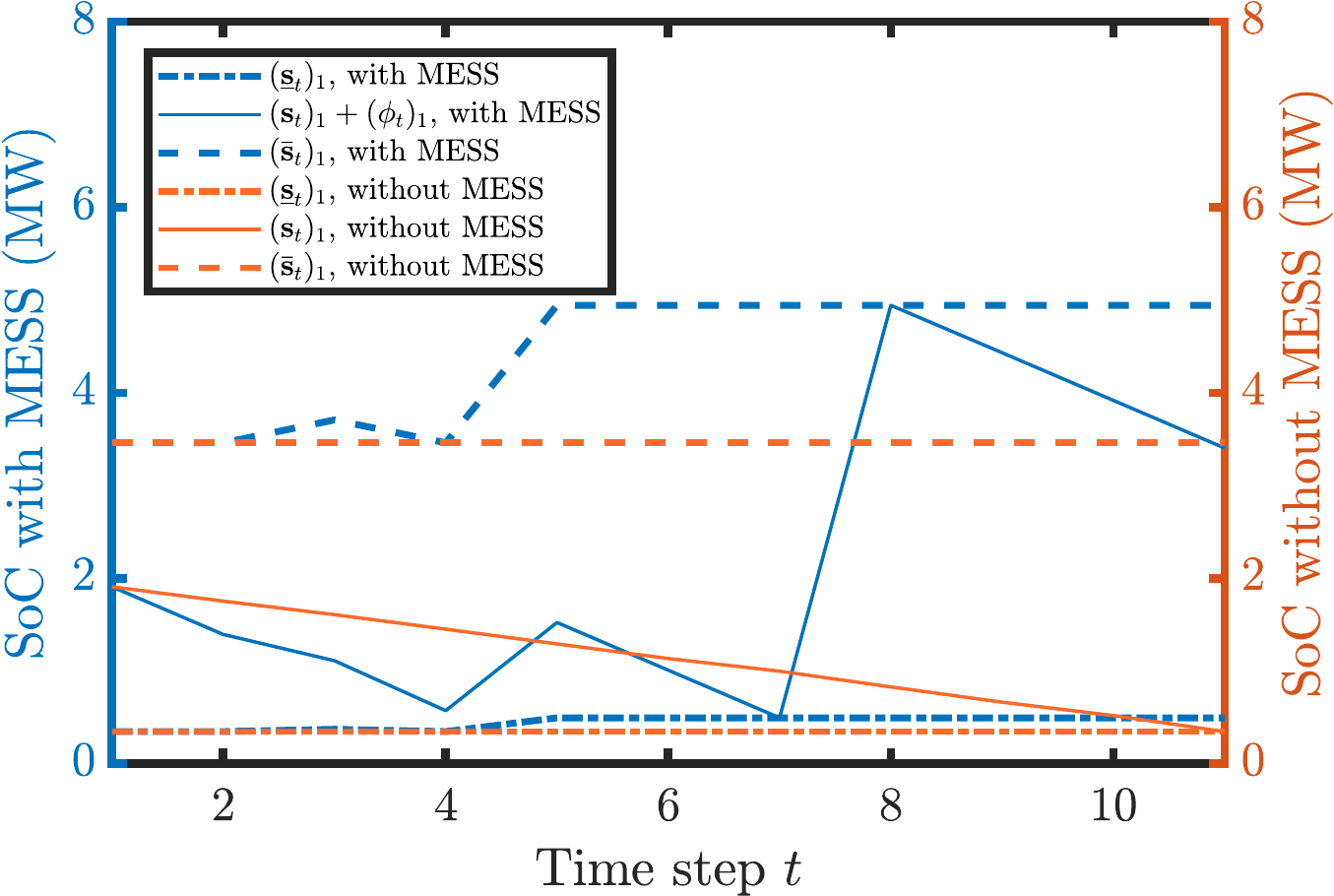}
		\caption{Comparison of evolution of SoC and SoC bounds at an ESS bus achieved with and without MESS.}
		\label{fig:soc_comparison}
	\end{figure}
	The LR problem, both with and without MESSs, is an MISOCP. We use Gurobi 9.3~\cite{gurobi} solver interfaced with Matlab via CVX~\cite{cvx}. Comparing the LR performance between the two methods in Fig.~\ref{fig:comparison_gen_load}, we see that in the case with no MESSs, the load power restored is significantly lower on all time steps than in the case with MESSs. This is a direct consequence of MESSs allowing SoC to reach buses in the MG where load to be restored is high as compared to generation and ESS discharge. Furthermore, we see that with MESSs, the generators produce more power in initial time steps, which is then stored in the MESSs and disbursed through all ESS buses. 
	Fig.~\ref{fig:soc_comparison} demonstrates the evolution of the SoC, as well as upper and lower bounds on SoC, for the cases with and without routing of MESSs. We see that with MESS routing, the upper and lower bounds on SoC evolve in time, thereby allowing the ESS to store more SoC than would be possible without MESSs. Finally, Table~\ref{fig:optimal_routing} lists the optimal routes of MESSs generated as the solution of the problem.
	
	\section{Conclusion} \label{conclusion}
	
	Considering the great flexibility of MESSs, our investigation of incorporating mobile battery units in MG operation is timely and inspiring. In this paper, we proposed a novel technique for load restoration in an MG equipped with MESSs. We presented the LR optimization problem, which co-optimizes the routing of MESSs with the restoration of loads. We demonstrated the efficacy of the proposed method over the counterpart with static ESSs through simulations. Topics of future studies involve increasing the scope of the problem to routing of any general resource in MGs, and exploitation of the MG network structure to reduce the problem complexity.
	
	\nocite{*}

\end{document}